\documentclass[traditabstract]{aa} 
\usepackage{graphicx}
\usepackage{colortbl}
\usepackage{rotating}
\usepackage{natbib}
\usepackage{sidecap}
\usepackage{txfonts}
\def \msun{\ifmmode{{\rm\ M}_\odot~}\else{${\rm\ M}_\odot$~}\fi}

\begin{document}
   \title{Steady jets from radiatively efficient hard states in GRS\,1915+105}



   \author{A. Rushton\thanks{ESO fellow}
          \inst{1}\inst{2}
          \and
          R. Spencer\inst{3}
	 \and
	  R. Fender\inst{4}\inst{5}
	  \and
	  G. Pooley\inst{6}
          }

   \institute{Onsala Space Observatory, SE-439 92 Onsala, Sweden
           \and
         European Southern Observatory, Karl-Schwarzschild-Str 2, 85748 Garching, Germany\\
         \email{Anthony.Rushton at eso.org}
         \and	   
         Jodrell Bank Centre for Astrophysics, School of Physics and Astronomy, University of Manchester, Manchester, M13 9PL\\
	      \email{Ralph.Spencer at Manchester.ac.uk}
	   \and
             School of Physics and Astronomy, University of Southampton, Southampton, SO17 1BJ\\
             \email{rpf at phys.soton.ac.uk}
	\and
	    Astronomical Institute `Anton Pannekoek', University of Amsterdam, Kruislaan 403, 1098 SJ Amsterdam, the Netherlands
	\and
	    The University of Cambridge, Mullard Radio Astronomy Observatory, Cavendish Laboratory, J. J. Thomson Avenue, Cambridge CB3 0HE\\
	    \email{guy at mrao.cam.ac.uk}
	}

   \date{Received 2010--05--04; accepted 2010--09--03}

 
  \abstract{

Recent studies of different X-ray binaries (XRBs) have shown a clear correlation between the radio and X-ray emission. We present evidence of a close relationship found between the radio and X-ray emission at different epochs for GRS\,1915+105, using observations from the Ryle Telescope and Rossi X-ray Timing Explorer satellite. The strongest correlation was found during the hard state (also known as the `plateau' state), where a steady AU-scale jet is known to exist. Both the radio and X-ray emission were found to decay from the start of most plateau states, with the radio emission decaying faster. An empirical relationship of $S_{\rm{radio }}\propto S_{\rm{X-ray}}^{\xi}$ was then fitted to data taken only during the plateau state, resulting in a power-law index of $\xi\sim1.7\pm0.3$, which is significantly higher than in other black hole XRBs in a similar state. An advection-flow model was then fitted to this relationship and compared to the universal XRB relationship as described by Gallo et al. (2003). We conclude that either (I) the accretion disk in this source is radiatively efficient, even during the continuous outflow of a compact jet, which could also suggest a universal turn-over from radiatively inefficient to efficient for all stellar-mass black holes at a critical mass accretion rate ($\dot{m}_{\rm{c}}\approx10^{18.5}$~g/s); or (II) the X-rays in the plateau state are dominated by emission from the base of the jet and not the accretion disk (e.g. via inverse Compton scattering from the outflow).
}
{}

   \keywords{Accretion, accretion disks --
               Black hole physics --
                X-ray: binaries --
                ISM: jets and outflows
               }

   \maketitle
%

\section{Introduction}

It has been well established that the formation of astrophysical jets is associated with accretion disks. Powerful jets have been observed from sizes that range from the AGN of extra-galactic quasars, to the much smaller ejecta around proto-planetary disks. Whilst the theory of accretion has been the focus of intense study for many decades, there exists no satisfactory explanation for this scale-free, invariant relationship between accretion disks and jets.

This paper presents a clear relationship between the accretion disk and jet formation in one of the most powerful black holes in the Galaxy, GRS\,1915+105. Using the high-resolution results and conclusions presented in \cite{2010MNRAS.401.2611R}, a confident assumption regarding the `state' of the jet can be made using radio and X-ray monitoring observations. As variations in the X-ray emission are associated with changes in the accretion disk, coupling these radio observations to simultaneous X-ray observations can directly relate to the inflow-outflow mechanisms around the black hole~\cite[reviewed by][]{2004ARA&A..42..317F}.

It has been clearly established that \textit{changes} in the X-ray spectral state of GRS\,1915+105 are associated with superluminal knots~\citep{1999MNRAS.304..865F}; however, large-scale structures quickly become divorced from the central accretion region as they move away and typically last for many days or weeks. To study the direct coupling between the accretion disk and jet, we must select observations of the compact `steady jet' when it is close to the energy source. As VLBI observations have shown this emission region to be only a few light hours in size~\citep{2000ApJ...543..373D}, changes in the accretion disk may therefore directly change the steady jet over the time-scale of hours. 

For the first time, observations taken over ten years have made it possible to clearly study the repetitive trends between the X-ray and radio. The results presented in this paper select radio and X-ray monitoring observations during the steady jet (`plateau') state. A direct relationship between the inflowing accretion and outflow jet can be made for GRS\,1915+105.

\subsection{X-ray spectral states of GRS\,1915+105}
\label{sec:X-ray_plateau}

The X-ray spectra of GRS 1915+105 can be generalised into two distinct components: a soft blackbody disk ($kT\sim1-2$~keV) and a hard non-thermal power-law extending to $\geq100$~keV associated with the very inner region or `corona' of the accretion disk. 

\cite{2000A&A...355..271B} initially identified 12 distinct X-ray classes with the RXTE-PCA based on the light-curve and colour-colour diagram of the source. From these classes, the overall X-ray spectra can be reduced as a transition between three basic states known as \textit{state A}, \textit{state B} and \textit{state C}; these states represent the interchange between the multi-temperature disk and the power-law component. States A and B correspond to a soft energy spectrum, with the dominant X-ray emission coming from the inner region of a thermal accretion disk with temperatures of $\sim1.8$~keV and $\sim2.2$~keV respectively. In-contrast, state C represents the near absence of this inner region and exhibits a dominant power-law component in the X-ray emission. 

\cite{2003A&A...404..283M} performed an analysis of the X-ray spectra with the RXTE-PCA, by isolating the observations into states A/B/C.  They fitted a blackbody component to the energy spectrum and directly related this to the inner radius ($R_{\rm{in}}=D\sqrt{N\cos\theta})$, where $D$ is the distance of the source, $\theta$ the inclination angle of the disk and N the normalisation of the thermal component. The inner radius was then compared to the overall temperature, $kT_{\rm{in}}$. It was found at the start of state C that $R_{\rm{in}}$ would rapidly increase in size and $kT_{\rm{in}}$ would drop to a much cooler temperature. This was explained as the collapse of the inner accretion region, leaving a truncated thin disk. During the evolution of the $R_{\rm{in}}-kT$ relationship in the state C, the temperature would slowly increase again and $R_{\rm{in}}$ would reduce. This has been suggested to be the re-filling of the inner region of the thin disk, until the return of the thermally dominant state of the XRB.

A detailed spectral observation of the state C by XMM-Newton was performed by \cite{2006A&A...448..677M}. They found only a simple power-law of $\Gamma\sim1.7$ was needed to fit to the overall spectra, which then gave residuals of a 1~keV excess, small variations between $\sim1.5-3$~keV and a deficit above 8~keV. The overall power-law was consistent with the RXTE-PCA observations, and was attributed to either a hot corona around the accretion disk or to Comptonized emission from the base of a jet \citep{2004ApJ...615..416R}. The 8~keV deficit was explained as an optically thick reflector that gave evidence of the presence of a thin disk. The 1~keV excess was (tentatively) explained as the presence of an optically thin component (e.g. a wind/jet or a geometrically thick disk). \cite{2006A&A...448..677M} noted that the relatively large amount of reflection components imply that the primary X-ray emitting region would have a size comparable to the inner disk radius.

\cite{2002MNRAS.331..745K} clearly established that radio emission is intimately related to the hard X-ray power-law component and implies a close physical connection between the inner accretion region and the outflowing synchrotron-emitting jet. They found a `one-to-one' relationship between radio oscillation events, originally discovered by \cite{1997MNRAS.292..925P}, and quasi-periodic X-ray dips during certain epochs in state C; in all other cases the source showed either `low-level' or `high-level' radio emission, but no radio oscillation \cite[see][for a detailed characterisation of the radio oscillation events]{2010ApJ...717.1222P}. Moreover, when the source stays in spectral hard state C for long periods of days to months, the `high-level' radio emission (i.e. non-oscillating emission) has been shown to originate from the compact flat-spectrum jet. This state is also known as the `plateau' state and is also referred to as class~$\chi$ by \cite{2000A&A...355..271B}.

\subsection{A universal X-ray binaries relationship and the fundamental plane}

X-ray binaries (XRBs) are ideal objects to study the disk-jet coupling relationship, as they display bright X-ray and radio emission from an accretion disk and jet, respectively. \cite{2000A&A...359..251C} found a clear relationship in the low-hard state of the XRB GX~339-4 between the X-ray and radio emission. The observed flat (or slightly inverted) spectrum was suggested to be a compact jet associated with the low-hard state. Discrete ejections of relativistic plasma are associated with transitional changes from this hard state to a softer state for most radio emitting black hole XRBs~\citep{2004MNRAS.355.1105F,2009MNRAS.396.1370F}. An empirical non-linear relationship between the radio and X-ray luminosity was found by \cite{2003A&A...400.1007C} as

\begin{equation}
L_{\rm{radio}}\propto L_{\rm{X-ray}}^{0.71\pm0.1},
\end{equation}

\noindent whilst the source remained in the low-hard state. \cite{2003MNRAS.344...60G}, then compiled a large sample of quasi-simultaneous radio and X-ray observations of stellar-mass black hole binaries, confirming this relationship to be universal for most black hole XRBs. 

Following this, \cite{2003MNRAS.345.1057M} and \cite{2004A&A...414..895F} independently linked the existence of a `fundamental plane' associated with all  accreting black holes, with the parameters $L_{\rm{radio}}$, $L_{\rm{X-ray}}$ and mass $M$. This relationship applied to a large range of masses, from super-massive black holes in extragalactic AGNs and the Galactic centre, Sgr~A$^{*}$, to stellar mass black holes in XRBs. This fundamental plane was found to take the form

\begin{equation}
L_{\rm{radio}}\propto L_{\rm{X-ray}}^{0.6}M^{0.8}.
\end{equation}
The aim of this paper is to test the fundamental relationship between the inflow and outflowing emission, with the source variability found in GRS\,1915+105. 

\section{Observations}

Data were obtained from instruments that have continually monitored GRS\,1915+105 to ensure no statistical biasing from particular periods activity or observational interest. The only radio telescope to have constantly observed GRS\,1915+105 over the last decade was the Ryle Telescope (RT). These data were then compared to $2-12$~keV X-ray data taken by the All Sky Monitor on board the RXTE satellite.

Data from each of the two instruments were binned into daily averages, and observations taken on the same day were cross-correlated to form a comprehensive radio-X-ray comparison. Whilst intra-day variability in both X-ray \citep{2000A&A...355..271B} and radio \citep{1997MNRAS.292..925P} is known to occur for various states, the aim of the work was not to study the details of a particular flare; rather, the goal was to study any longer term correlation (i.e. longer than a few days) between the radio and X-ray in the steady jet state. 

Moreover, GRS\,1915+105 does not always produce compact self-absorbed emission and large-scale, optically thin structure is frequently observed after a state change. Only when a compact jet, with no superluminal proper motion is present, can a direct comparison between the radio and X-ray be made. Therefore, it was important to identify the states that were not contaminated by large scale extended knots, such as those observed by~\cite{1994Natur.371...46M}. 

\subsection{Ryle Telescope}

The RT is a linear east-west radio interferometer located at the Mullard Radio Astronomy Observatory, UK. The array operates at a frequency of 15~GHz with associated baselines between approximately 18~metres and 4.8~kilometres, although for the majority of these observations only a subset of the baselines, typically up to 150~metres, were used.

An extensive monitoring campaign began with the RT in 1996 of a few radio-bright X-ray binaries (including GRS\,1915+105), which coincided with the launch of the RXTE satellite. Observations of target sources were interleaved with a nearby phase calibrator (B1920+154 in the case of GRS\,1915+105) and the flux-density scale was set by short scans of 3C~48 and 3C~286. The data were sampled every eight seconds and averaged into five minute data points with an RMS of $\sim2$~mJy.

Near daily observations of GRS\,1915+105 were collected between May 1995 (MJD 49856) and June 2006 (MJD 53898) at 15 GHz. \cite{1997MNRAS.292..925P} describe the details of this programme, indicating the detection of a 20 -- 40 minute quasi-periodic variation of GRS\,1915+105 at a frequency of 15 GHz, apparently coupled with variations in the soft X-rays~\citep[see][for a comprehensive study]{2010ApJ...717.1222P}. 

\subsection{Rossi X-ray Timing Explorer satellite}

The All Sky Monitor (ASM) instrument on-board the RXTE has been monitoring the sky since March 1996 and the data presented here covers the period from March 1997 to 2007. With each orbit of the RXTE, the ASM surveyed $\sim80\%$ of the sky to a depth of $20-100$~mCrab, making approximately ten observations of a source per day. A more detailed description of the RXTE-ASM can be found in~\cite{1996ApJ...469L..33L}.

Each individual pointing, or dwell, was a 90 second integration of the source, with intensities measured in three energy bands of $1.5-3$, $3-5$, and $5-12$~keV. The Crab Nebula flux between $1.5-12$~keV corresponds to about 75 ASM counts~s$^{-1}$. To calculate the spectral hardness ratio (HR2), individual dwells were averaged into daily points and the ratios between the $5-12$~keV and $1.5-3$~keV energy bands were taken. 

\section{Results and analysis}

%
%
%

\begin{SCfigure*}
\centering
\includegraphics[width=12cm, angle=0, trim=0 -10 0 0]{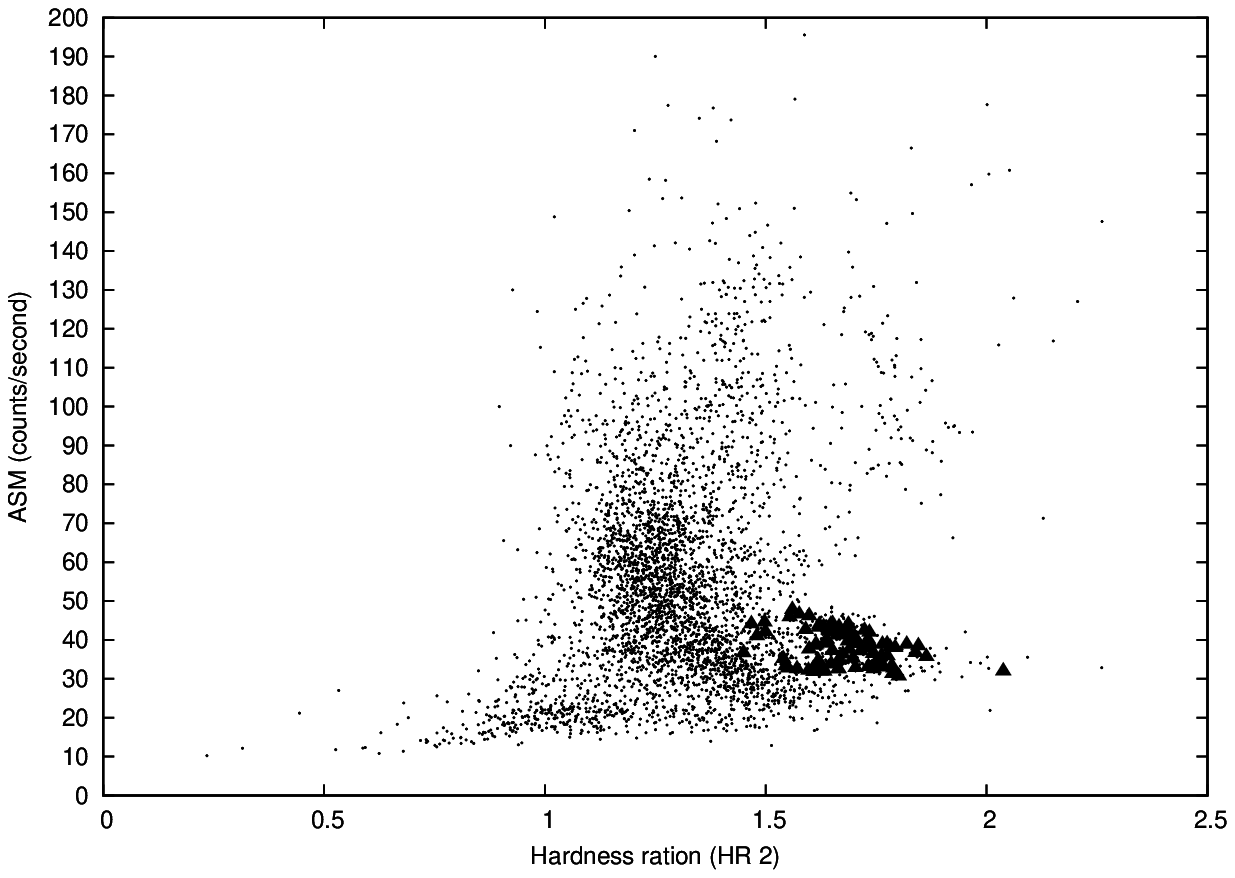}
\caption{\label{fig:HID}Hardness intensity diagram of GRS\,1915+105 between 1996 and 2008, showing the one day average intensity, between 2-12~keV, plotted as a function of the hardness radio, HR2 ($\frac{5-12\rm{~keV}}{2-5\rm{~keV}}$). The triangles represent \textit{persistent radio emission} associated with the X-rays known as class~$\chi$ by \cite{2000A&A...355..271B} and the `plateau' state by \cite{1997MNRAS.292..925P}.}
\end{SCfigure*}

\subsection{Absorption correction}

The observed X-ray flux density from the RXTE-ASM, measured in counts per second, is partially absorbed by hydrogen within the line-of-sight. The data were therefore compensated for the level of absorption using previous high-spectral resolution observations. \cite{2003MNRAS.344...60G} modelled the spectra of hard state BHs, with hydrogen column densities ($N_{\rm{H}}$) of zero up to $12.5\times10^{22}\rm{cm}$ using Chandra observations. By fixing the flux corresponding to no absorption, it is possible to model the data points using a simple exponential function: 

\begin{equation}
\frac{F_{\rm{abs.}}}{F_{\rm{unabs.}}}=\exp\left[\frac{-(N_{\rm{H}}/10^{22}~\rm{cm}^{-2})}{18.38}\right].
\end{equation}
We then used a line-of-sight hydrogen column density for GRS\,1915+105 of $N_{\rm{H}}=5\times10^{22}~\rm{cm}$ as found by \cite{1994AIPC..304..260G} .
\rm
\subsection{Hardness Intensity Diagram}

To identify the different X-ray spectral states, the RXTE-ASM data were sorted by comparing the spectral hardness ratio, HR2 ($\frac{5-12~\rm{keV}}{2-5~\rm{keV}}$), with the  total X-ray intensity ($2-12$~keV). This is shown in Figure~\ref{fig:HID} as a hardness intensity diagram (HID).

The diagram in Figure~\ref{fig:HID} can broadly be divided into three dominant modes of X-ray emission:
\begin{itemize}
\item \textbf{Weak/very soft} flux of $<20$~c/s with a HR2~$<1$
\item \textbf{Highly variable/soft} flux between $20-200$~c/s with a soft HR2~$\approx1-1.5$
\item \textbf{Persistent/slightly harder} flux at approximately $30-50$~c/s with a relatively harder spectra of HR2~$\approx1.5-2$
\end{itemize}

\noindent These are akin to the X-ray states B, A and C respectively, described by \cite{2003A&A...404..283M}. Resolved jets are observed when the source transitions from state C to A~\citep{1999MNRAS.304..865F,2010MNRAS.401.2611R}.

Next, the X-ray emission associated with little or no radio flux was identified. This allowed areas of the HID that showed no radio correlation to be discounted. It was found that during the softer X-ray states of HR2$<1.5$), there was no associated radio emission that lasted more than one or two days. Likewise, the strong X-ray flares of $>50$ counts/second were not associated with any long periods of strong radio emission. A substantial fraction of the X-ray emission in the HID is therefore not associated with any strong radio flux and demonstrates no clear connection between the two emission mechanisms.

\label{sec:plateau-selection}

The clearest correlation between radio and X-ray flux observed in GRS\,1915+105 was during the persistent/slightly harder X-ray state (as marked as triangles in Figure~\ref{fig:HID}). When the X-ray flux is persistently (i.e. more than a few days) between $30-50$~counts/second and has a hardness radio of HR2~$>1.5$, the source is \textit{always associated with strong radio flux}. This state was first identified by \cite{1997MNRAS.292..925P} as the `plateau' state and later by \cite{2000A&A...355..271B} with the RXTE-PCA as class~$\chi$. It is also the state where a compact jet is always present \citep{2000ApJ...543..373D,2003A&A...409L..35F}.

\subsection{Radio/X-ray relationship}

\begin{figure*}
  \begin{center}
\includegraphics[width=18cm, angle=0, trim=0 -10 0 0]{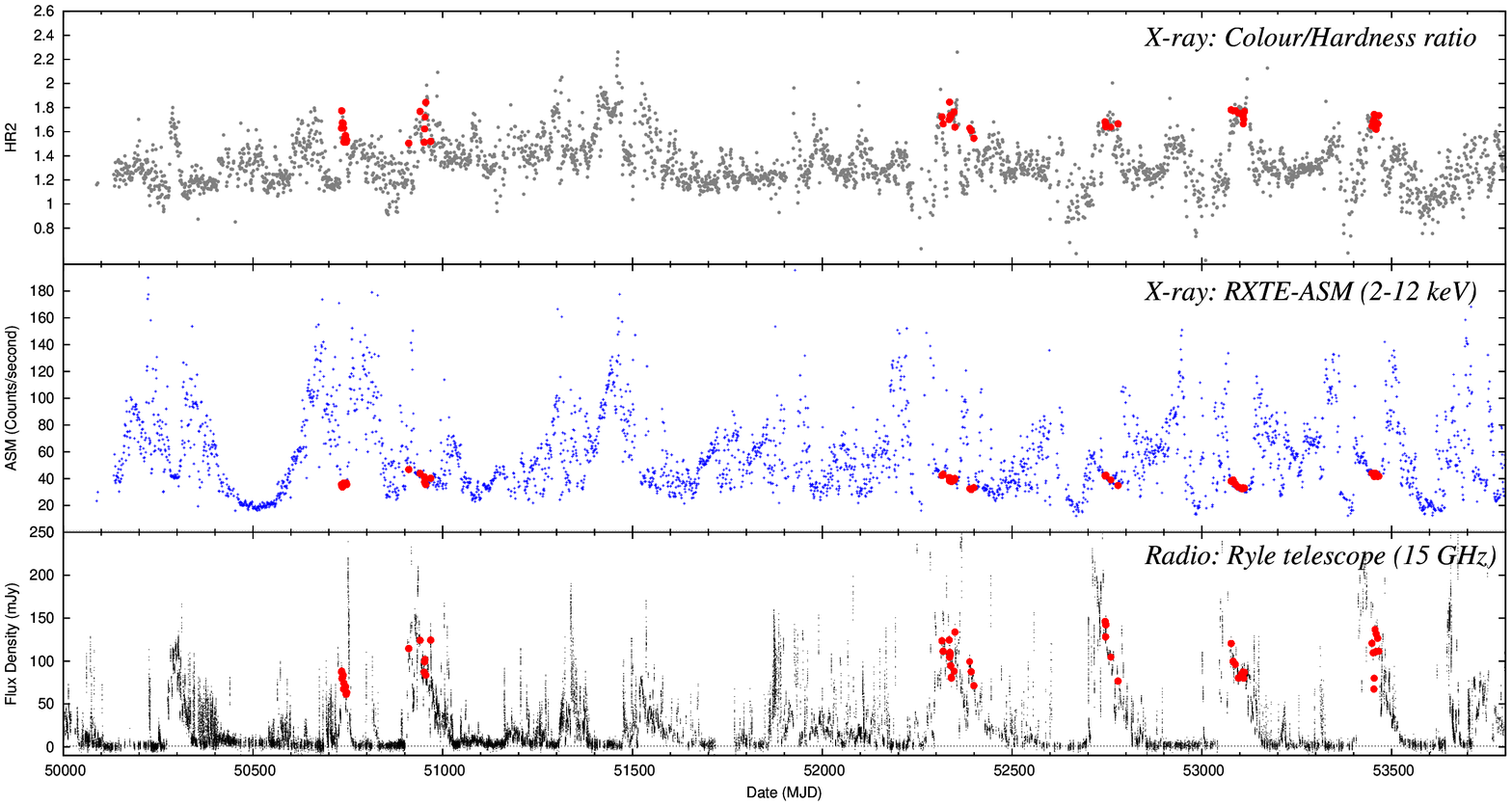}
\end{center}
\caption{\label{fig:1915-lightcurve}X-ray and radio lightcurves of GRS\,1915+105 over a ten year period with the red marks representing \textit{persistent radio emission} associated with class~$\chi$ or the `plateau' state (i.e. ASM counts of $30-50$ per second and HR2~$>1.5$); the bottom graph shows 15 GHz radio observations taken by the Ryle Telescope, the middle and top graphs show $2-12$ keV X-ray and HR2 hardness ratio ($\frac{5-12~\rm{keV}}{2-5~\rm{keV}}$) respectively observations taken with the RXTE-ASM.}
\end{figure*}

\begin{table*}
\begin{center}\begin{tabular}{|>{\columncolor[rgb]{0.8,0.8,0.8}}l>{\columncolor[rgb]{0.8,0.8,0.8}}l|ccc|}
\hline
\rowcolor[rgb]{0.8,0.8,0.8} Start date & End date & Change in X-rays$^{\dagger}$ & Change in radio$^{\ddagger}$ & Duration \\
\rowcolor[rgb]{0.8,0.8,0.8}   (MJD)   & (MJD) &  (counts/second)       &   (mJy) & (Days) \\
\hline
50286 & 50297 & $42 \rightarrow 40$ & $110 \rightarrow 100$ & 11 \\
50732 & 50736 & $36 \rightarrow 33$ & $94 \rightarrow 87 $ & 4 \\
50926 & 50989 & $45 \rightarrow 34$ & $154 \rightarrow 31$ & 63 \\
51522 & 51546 & $39 \rightarrow 33$ & $78 \rightarrow 50 $ & 24 \\
52315 & 52354 & $43 \rightarrow 36$ & $ 122 \rightarrow 73$ & 41 \\
52729 & 52783 & $46 \rightarrow 33$ & $ 163 \rightarrow 68$ & 54 \\
53077 & 53116 & $39 \rightarrow 31$ & $ 122 \rightarrow 70$ & 39 \\
53416 & 53466 & $52 \rightarrow 42$ & $ 197 \rightarrow 112$ & 50 \\
\hline 

\end{tabular}\end{center}
\caption{List of the hard states (i.e. `plateau' state/class~$\chi$) of GRS\,1915+105 between January 1996 (MJD~50110) and May 2006 (MJD~53883). The X-ray observations were measured as one day averages of the RXTE-ASM observations between 2-12~keV. Radio observations were taken at 15 GHz with the Ryle telescope (RT). $\dagger$ RXTE-ASM errors were typically less than 1 count/second. $\ddagger$ RT errors are estimated at approximately $\pm6$~mJy.}
\label{table:RXTE-ASM_RT}
\end{table*}

The X-ray and radio lightcurves of GRS\,1915+105 over the ten year period between 1996 and 2006 are shown in Figure~\ref{fig:1915-lightcurve}. Marked red are the points selected in the persistent/slightly harder X-ray state (i.e. the compact jet state identify as triangles in Figure~\ref{fig:HID}). It is clear that the long flaring periods of radio activity are coincident with this X-ray state. Furthermore, both the X-ray and radio emission appear to decay during each flaring period. Table~\ref{table:RXTE-ASM_RT} lists eight periods of flaring activity (ranging from 4 to 63 days) identified between January 1996 (MJD~50110) and May 2006 (MJD~53883). It thus appears that the `plateau' state occurs approximately every 1.3 years.

\begin{figure*}
  \begin{center}
\includegraphics[width=18cm, angle=0, trim=0 -10 0 0]{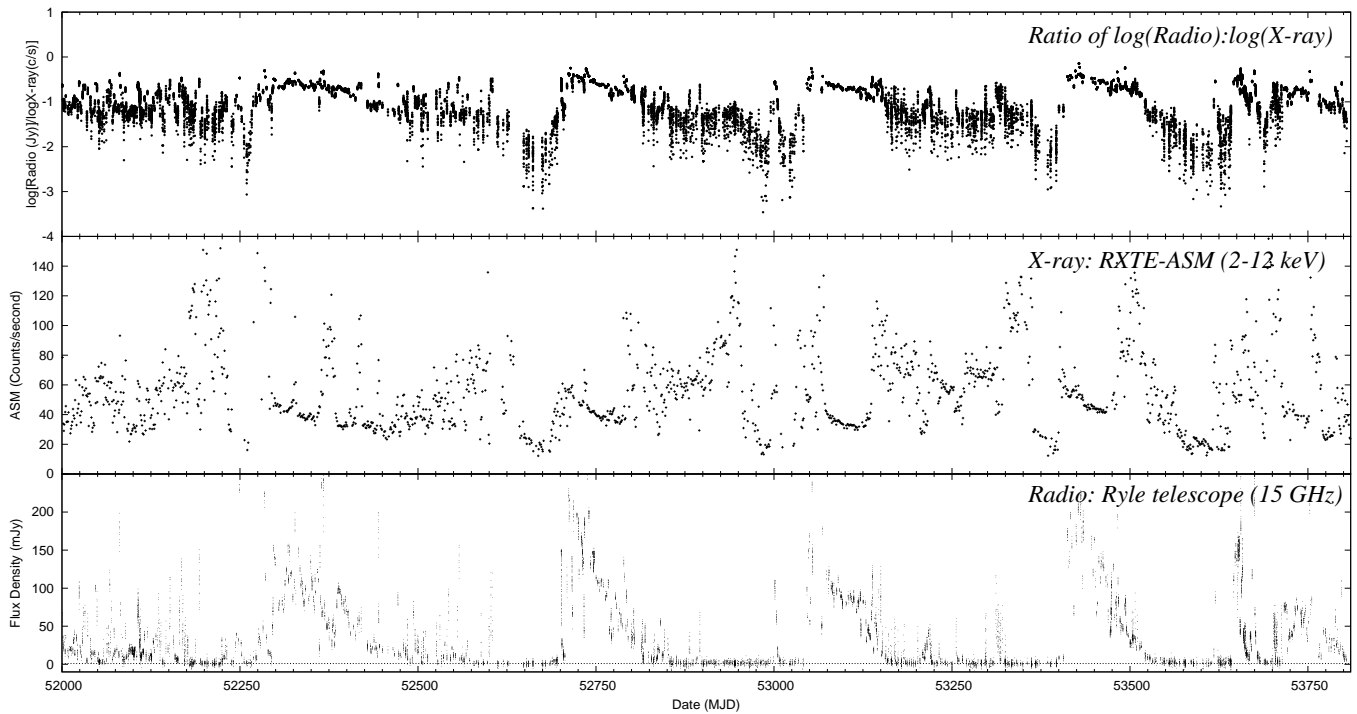}
\end{center}
\caption{\label{fig:ratio}The radio and X-ray lightcurves (bottom and middle respectively) of five long periods of outburst. The top graph represents the ratio of $\log(\rm{radio}) / \log(\rm{X-ray})$, clearly showing direct relationship between the two bands.}
\end{figure*}

Furthermore, the radio/X-ray correlation appears particularly strong during a period between April 2001 (MJD~52000) and February 2006 (MJD~53767), as shown in Figure~\ref{fig:ratio}. Five periods of strong radio outburst are clearly identified separated by approximately one year. During each period both the radio and X-ray emission appear to decay from the beginning of each outburst. As the relationship between radio and X-ray has been shown as non-linear for XRBs~\citep{2003MNRAS.344...60G}, the ratio of log(radio)~:~log(X-ray) is also shown in Figure~\ref{fig:ratio}~(top). The data shows a strong coupling between the two wave-bands during the plateau state and there is also tentative evidence that the coupling extends beyond this state, but the latter result is not conclusive; however, it is clear that the radio emission decays more quickly than the X-ray from the start of most plateau states.

The ratio between the bands appears quasi-periodic over outbursts in a saw-tooth pattern: there is a short period of radio quenching lasting 10-20 days, followed by the plateau hard state lasting $\sim100$~days where the radio decreases more rapidly than the X-ray emission, a period of~$\sim200$~days where the radio is weaker and the X-ray are highly variable, returning to the quenched state again.  

\section{Accretion and jet formation models}
\label{section:accretion}

Understanding the processes connecting in-falling matter with outflowing jets requires measuring the mass accretion rate ($\dot{m}$), which fundamentally connects the mechanisms responsible for radio and X-ray emission. The following derives a relationship assuming an underlining geometry and physics, based on the accretion process and power of the jet.

Accretion disks can be either an efficient or inefficient mechanism for lowering material into a gravitational potential and extracting energy as radiation. The vital process is converting orbital kinetic energy into heat. As the accretion around a black hole inevitably involves rotating gas flows, we must considered hydrodynamic equations of viscous differentially rotating flows. \cite{1995MNRAS.275..641C} describes four known self-consistent solutions, two of which (a thin disk an advection-dominated flow, as shown in Figure~\ref{fig:schematic}) are considered in Sections~\ref{section:thin disk} and \ref{section:adaf}.

\subsection{Thin disk}\label{section:thin disk}

In many situations the accretion flow onto a compact object can be approximated by a two-dimensional gas flow. A thin disk model around a compact object can be used to model the thermal emission from a binary system~\citep{1973A&A....24..337S}. The accreting gas is assumed to form a geometrically thin, optically thick disk, producing a blackbody spectrum. The X-ray spectra of XRBs in the soft state, have been successfully modelled as the thermal emission from the inner accreting region of such a thin disk.

In a binary system the outflowing matter from a normal star will have much higher angular momentum than the compact object. The particles in the disk will lose angular momentum due to interactions between the various layers. A compact object with mass $M$ will then accrete with a luminosity (at an efficiency $\eta$) of

\begin{equation}
L_{\rm{disk}}=2\eta\frac{GM\dot{m}}{R_*},
\end{equation}
when the matter has fallen to radius $R_*$~\citep{2002apa..book.....F}. Such an accretion process is therefore considered radiatively efficient and one can estimate $\dot{m}$ directly from the X-ray bolometric luminosity ($L_{\rm{X}}$). Assuming a disk efficiency $\eta$, the accreting rate for luminosity is given by

\begin{equation}
\dot{m}=\frac{L_{\rm{X}}}{f\eta c^2}=1.5\times10^{18}\left(\frac{L_{\rm{X}}}{10^{38}\rm{erg~s}^{-1}}\right)\left(\frac{0.75}{f}\right)\left(\frac{0.1}{\eta}\right)\rm{\frac{g}{s}},
\label{eq:thin_disk_rate}
\end{equation}

\noindent where $f$ is the fraction of the total accretion that is not ejected via jets or winds.

\subsection{Advection-dominated flows}\label{section:adaf}

Radiatively efficient thin-disk models have successfully described the basic accretion properties of a thermally-dominant disk; however, this disk model has only been suggested over a certain range of accretion rates and breaks down when the gas pressure is below a critical value. In recent years, alternative solutions for radiatively inefficient accretion flows have therefore become popular mechanisms for explaining the low-hard/quiescent state in stellar black holes \citep[reviewed by][]{2006csxs.book..157M}. The inflow of plasma forms a geometrically thick disk and the rate of gas cooling is such that most of the dissipated energy is not radiated.

Within advection-dominated accretion flow~\citep[ADAF;][]{1982Natur.295...17R,1995ApJ...438L..37A} the accreted gas is held in a low density quasi-spherical `corona' around the compact object. The lower accretion rates of the thick disk inhibit Coulomb coupling between electrons and ions, trapping part of the viscous energy as heat within the gas. Transport of angular momentum then occurs as the viscous energy is ``advected'' onto the black hole, rather than being radiated away. Without the viscosity, no accretion would occur, as the angular momentum would prevent direct radial in-fall. For black holes with no physical surface,  part of the advected material will cross the event horizon and not escape. This results in some of the accreting power becoming lost and making the ADAF radiatively inefficient (i.e. the bolometric luminosity will be lower than expected from a given mass accretion rate). This is perhaps evidence of the existence of an event horizon, and predicts an X-ray luminosity of black holes that scales as

\begin{equation}\label{eq:ADAF_accretion-rate}
L_{\rm{X-ray}}\propto \dot{m}^2.
\end{equation}

\begin{figure}
  \begin{center}
\includegraphics[width=9cm]{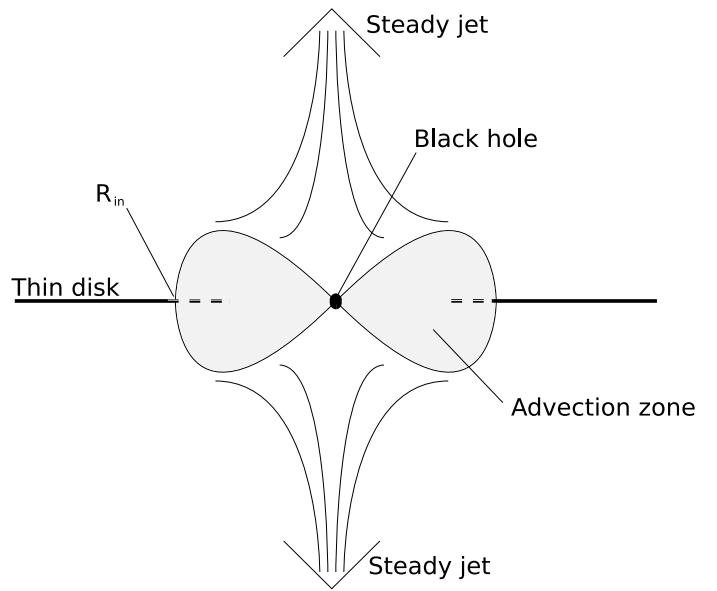}
\end{center}
\caption{\label{fig:schematic}A schematic drawing of advection accretion. Accretion onto a black hole forms a thin disk until a critical radius, $R_{\rm{in}}$, where upon the accretion flows in geometrically thick disk, known as an advection flow. Powerful bipolar jets are formed in the advection zone and are sensitive to the mass accretion, $\dot{m}$ and the size of the advection zone, $R_{\rm{in}}$.}
\end{figure}

\subsection{Jet power}

The precise details of how a scale-free relativistic jet can be formed from an accretion disk remain unknown; however, it is possible to calculate the total power of a partially absorbed jet and can be used as a `tracer' for the mass accretion. Equation 56 in \cite{1995A&A...293..665F} shows under the same constraints, the observed luminosity of the self-absorbed jet depends on the power as

\begin{equation}
L_{\rm{Radio}}\propto Q_{\rm{jet}}^{17/12},
\label{eq:jet_power}
\end{equation}

\noindent if modelled by a simple conical jet~\cite[a similar result was predicted by][]{1979ApJ...232...34B}. Assuming that the jet forms a linear inter-dependency with the accretion disk, the power of the jet will then be a constant fraction of the outflow rate $Q_{\rm{jet}}=q\dot{m}c^2$, with an efficiency $q=10^{-3}-10^{-1}$~\citep{1995A&A...293..665F}. We therefore find

\begin{equation}\label{eq:radio_accretion-rate}
\dot{m}=\dot{m}_0\left(\frac{ L_{\rm{Radio}}}{L_{\rm{Radio,0}}} \right)^{12/17},
\end{equation}
where $\dot{m}_{0}$ and $L_{\rm{Radio,0}}$ are normalization factors (assuming a flat spectral index of $\alpha=0$ and an energy distribution of $\rm{d}N_e\propto E^{-p}$ where $p=2$). \cite{2006MNRAS.369.1451K} then showed that by assuming the accretion rate does not vary much during a state change, it is possible to estimate $\dot{m}_0$ from the soft state X-ray luminosity (i.e. using Equation~\ref{eq:thin_disk_rate}). They found that for a sample of neutron star~(NS) and black hole~(BH) sources, with known accretion rates that 
\begin{equation}
\dot{m}_0^{ \rm{NS} }=7.7\times10^{17}$ \rm{g/s}$~~~$\rm{and}$~~~\dot{m}_0^{\rm{BH}}=4.0\times10^{17}~\rm{g/s},
\end{equation}
when setting $L_{\rm{Radio,0}}=10^{30}$~ergs as this is the radio luminosity where the accretion disk around a 10\msun black hole changes it state.

\section{Model fitting}

\begin{SCfigure*}
    \centering
\includegraphics[width=12cm]{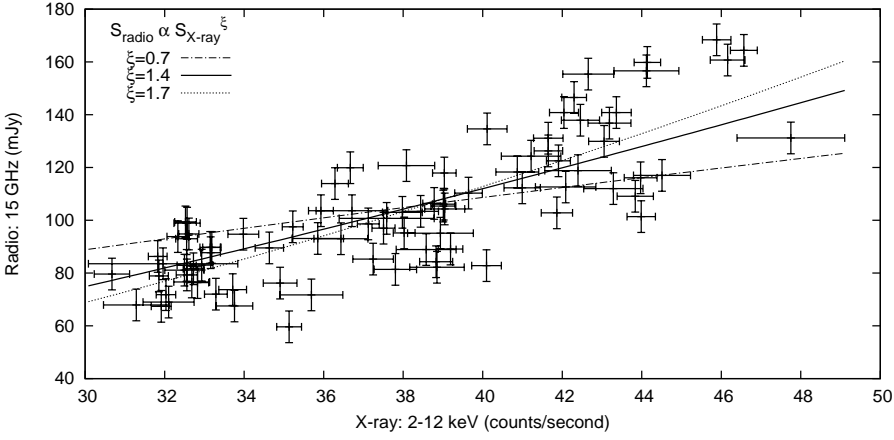}

\caption{\label{Radio_X-ray}X-ray/radio correlation in the hard state of GRS\,1915+1915 (i.e. plateau state). The dotted line represents a `best-fit' of $\xi\sim1.7\pm0.3$, the solid line represent a model of $\xi\sim1.4$ and the dashed-dotted line represents a model of $\xi\sim0.7$, where $S_{\rm{radio}}\propto S_{\rm{X-ray}}^{\xi}$.}
\end{SCfigure*}

The data selected in Section~\ref{sec:plateau-selection} are plotted in Figure~\ref{Radio_X-ray} to search for an empirical relationship between the X-rays and radio. To determine if any statistical relationship existed, a Spearman's rank correlation test \citep{1989sgtu.book.....B} was used to quantify the level of correlation between the X-rays and radio, without assuming any parametrized model. The Spearman's rank correlation coefficient was found to be 0.81 (i.e. statistically significant) and the analysis thus clearly shows a strong relationship between the X-rays and radio. This indicates a close coupling between the mechanisms producing the radio and X-rays whilst GRS\,1915+105 is in the hard state.

We then fitted a function of $S_{\rm{radio}}\propto S_{\rm{X-ray}}^{\xi}$, with an index of $\xi\sim1.7\pm0.3$, which is much steeper than the previous correlations that have shown $\xi\sim0.7$ for XRBs in the low-hard state. Previous models have then assumed that the bolometric luminosity is dominated by an advection flow and thus the observed X-ray emission is $L_{\rm{X-ray}}\propto \dot{m}^2$ (i.e. Equation~\ref{eq:ADAF_accretion-rate}). As the radio emission relates to the accretion rate as $ L_{\rm{radio}}\propto \dot{m}^{17/12}$ (Equation~\ref{eq:radio_accretion-rate}), the radio emission should couple to the X-ray emission as

\begin{eqnarray}
L_{\rm{rad}}\propto L_{\rm{X}}^{0.7},
\label{eq:inefficient-accretion}
\end{eqnarray}
which is therefore a radiatively inefficient coupling. However, the fit has a much steeper slope, that can be better fitted without assuming an advection flow. If the X-ray emission coupled linearly, $L_{\rm{X-ray}}\propto \dot{m}$ (Equation~\ref{eq:thin_disk_rate}), with the accretion rate then

\begin{eqnarray}
L_{\rm{rad}}\propto L_{\rm{X}}^{1.4},
\label{eq:efficient-accretion}
\end{eqnarray}
which is therefore a radiatively efficient coupling and gives a much closer fit to the observed relationship.

\begin{SCfigure*}
   \centering
\includegraphics[width=12cm]{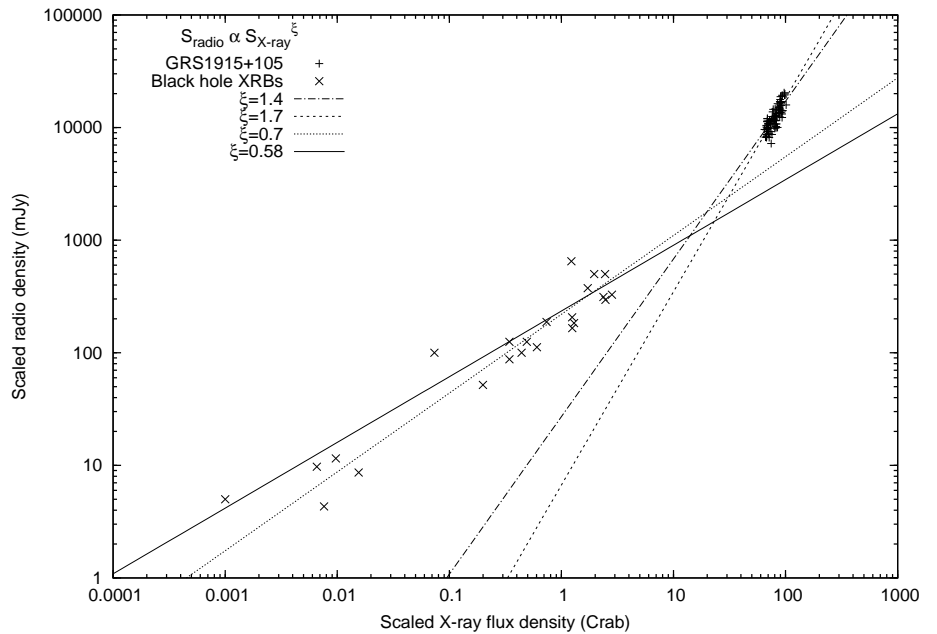}
\caption{\label{fig:universal_plane}Scaled X-ray/radio correlation in the hard state (i.e. plateau state) of GRS\,1915+1915 and other known black hole XRBs. Each source has been scaled to a distance of 1~kpc. The dashed-dotted and dotted lines represent the efficient and inefficient models of $\xi\sim1.4$ and $\xi\sim0.7$, respectively, where $S_{\rm{radio}}\propto S_{\rm{X-ray}}^{\xi}$. The solid and dashed lines represent an empirical `best-fit' to each XRB.}
\end{SCfigure*}

This trend can be compared to other XRBs also whilst in the low-hard state \citep[e.g. ][]{2003MNRAS.344...60G,2006MNRAS.370.1351G}. This showed a similar power-law relationship between X-rays and radio, but with a much steeper slope. GRS\,1915+105 and the other XRB luminosities were then scaled to 1~kpc and assumed to have approximately the same black hole mass. The X-ray/radio correlation was therefore compared for two similar black hole accretors, in a similar X-ray spectral state, as shown in Figure~\ref{fig:universal_plane} and in more detail for GRS1915+105 in Figure~\ref{Radio_X-ray}. Note that correlation in Figure~\ref{Radio_X-ray} is equivalent to the smooth part of the log (radio) / log (X-ray)  plots in Figure~\ref{fig:ratio} during the hard state (state C). The dashed-dotted and dotted lines, represent the different models: radiatively \textit{efficient} and radiatively \textit{inefficient}, as in Equations~\ref{eq:inefficient-accretion} and \ref{eq:efficient-accretion}. The solid and dashed lines represent the empirical `best-fits' to the data. It is therefore suggested that despite the similarities in the fundamental parameters of the sources, the radiative mechanisms scaling the bolometric luminosities of GRS\,1915+105, to that of the \cite{2003MNRAS.344...60G} black hole sources, is physically different.

\subsection{Estimates of the accretion rate}

The bolometric luminosity of a XRB in the low/hard state does not give a reliable estimate of the mass accretion. It was therefore suggested by \cite{2006MNRAS.369.1451K} that the radio emission of XRBs, with known accretion rates, can be used to scale Equation~\ref{eq:radio_accretion-rate}. For GRS\,1915+105 this has been found as $\dot{m}\sim 10^{19}~$g/s. Using the scaling relationship, $ L_{\rm{radio}}\propto \dot{m}^{17/12}$, this accretion rate was extrapolated to the lower radio luminosities observed in the other black hole XRBs.

\begin{SCfigure*}
   \centering
\includegraphics[width=12cm]{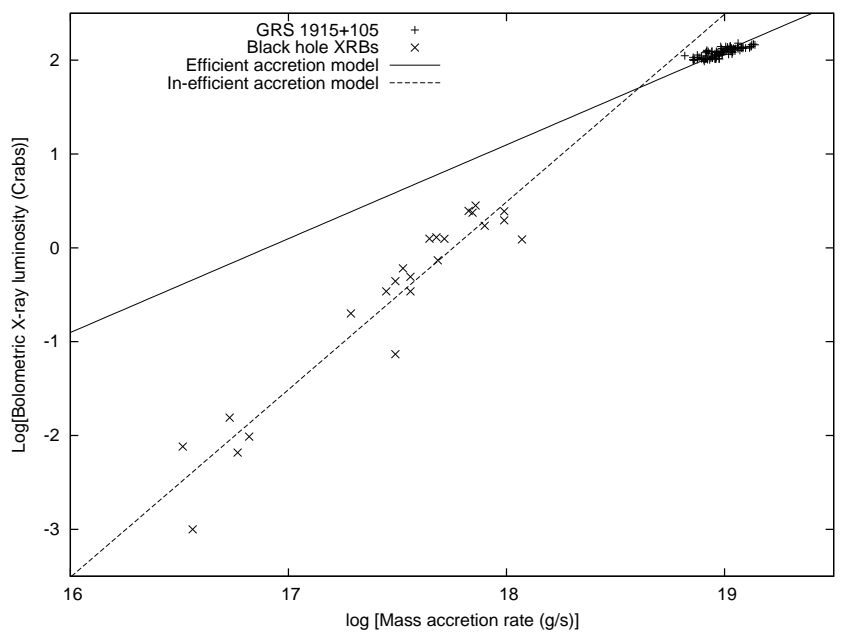}
\caption[The bolometric luminosity of known black hole XRBs and GRS\,1915+105 as a function of accretion rate $\dot{m}$.]{\label{fig:accretion-rates}The bolometric luminosity of known black hole XRBs and GRS\,1915+105 as a function of accretion rate $\dot{m}$. The solid line represents the predicted X-ray bolometric luminosity of a radiatively efficient XRB. The dashed line represents the expected X-ray bolometric luminosity for different rates of advection flow accretion.}
\end{SCfigure*}

The estimated accretion rate was then compared to the bolometric X-ray luminosity, as shown in Figure~\ref{fig:accretion-rates}. The solid line represents the predicted X-ray bolometric luminosity of a radiatively efficient XRB. The dashed line represents the expected X-ray bolometric luminosity for different rates of advection flow accretion. As shown on the graph, most black hole XRBs are far below the expected X-ray luminosity if all of the heat energy released from angular momentum loss was radiated away. However, variations in GRS\,1915+105 do not show this relationship; instead, there appears to be a `turn-over' around $\dot{m}\sim10^{18.5}$~g/s.

\section{Discussion}

It is apparent from the results presented here, that the X-ray/radio variations between GRS\,1915+105 and other black hole XRBs do not exhibit the same scaling relationship. Two possible explanations are therefore suggested:

\subsection{Radiatively efficient accretion} 

The evidence presented by \cite{2003A&A...404..283M} suggests that during state C (including the steady jet state) there is a loss of matter in the most inner region ($r_{\rm{in}}$) of an optically thick and geometrically thin disk. The collapse of such an inner region is believed to be due to a reduction in the mass accretion rate, thus forming an advection dominated region (as shown in Figure~\ref{fig:schematic}), that can produce a steady outflow of high-energy particles; however, the X-ray emission is a factor of $\sim100$ brighter in GRS\,1915+105 (and relatively softer) than most XRBs observed in the low/hard state. Furthermore, XMM-Newton observations \citep{2006A&A...448..677M} suggest the presence of an optically thick reflector, that provides evidence of the presence of a thin disk in the X-ray spectrum. 

It is therefore suggested that the domiante mode of accretion is efficient accretion, where the X-ray emission is emitted by a \textit{radiatively efficient thin disk}, where $L_{\rm{X}}\propto\dot{m}$, rather than an advection dominated flow. This linear dependence comes from the properties of a surrounding corona, rather than the thin disk and may only be valid for specific geometries and physical conditions. One must assume that the fraction of gravitational energy~($A$) dissipated from the thin disk into the corona is constant and the power of the corona is $Q_c=A \eta \dot{m}c^2$. The energy stored in the corona can then be radiated away by inverse Compton scattering of the disk photons such that $L_{X} \propto \dot{m}$, before reaching $r_{\rm{in}}$. 

These argument have profound consequences on the physical conditions required to form an outflowing jet; there exists a rate which the nature of accretion fundamentally changes. Figure~\ref{fig:accretion-rates} suggests, for stellar-mass black holes, the dominant mode of accretion changes from radiatively inefficient to radiatively efficient around a mass-accretion rate of $\dot{m}\sim10^{18.5}$~g/s. This would imply that at high $\dot{m}$ that advection flow accretion does not dominate the bolometric luminosity, despite strong evidence of a continuous outflow of particles (i.e. the compact jet). Furthermore, this argument can be scaled with the mass of the black hole, meaning advection dominated flows are not sustainable at a very high fraction of the Eddington accretion rate of about $\dot{m}>0.1~\dot{m}_{\rm{edd}}$. It is therefore speculated that the change in accretion mode is a fundamental physical invariant that could occur from stellar-mass black holes to supermassive black holes in AGN.

\subsection{Jet dominated emission}

At very high-mass accretion rates, there could be a turn-over in the dominance of the X-ray bolometric luminosity between the accretion disk and the jet. The close relationship between X-rays and radio emission shown in Figure~\ref{fig:ratio} and Figure~\ref{Radio_X-ray} could be explained if the respective radiation fields originate from the same outflowing particles \citep[as suggested for XTE J1118+480 by][]{2001A&A...372L..25M}. This argument is further supported by \cite{2008ApJ...675.1449R}, who used INTEGRAL and RXTE observations of GRS\,1915+105, during class $\chi$/plateau state, to model the X-ray spectra into two components: a thermal comptonized corona and an additional power-law tail between $2-200$ keV. They found a direct correlation between variability in the hard tail and the strong radio emission from the compact jet (using the RT at 15~GHz). The X-ray bolometric luminosity may therefore not be dominated by the accretion disk and the underlying radiative efficiency of the disk (and hence mass accretion rate) would not significantly contribute to the bolometric luminosity; rather the overall X-ray luminosity would simply become some function of the outflow rate.

Representative parameters for the radio jet in the long hard state can be found from \cite{2000ApJ...543..373D} where their epoch~E showed a jet $5~\times~(<1)$~mas$^2$ with a flux density of 64 mJy. Assuming a spectral index of 0.4 and limits to the radio regime of 10~MHz to 100~GHz we find a total radio luminosity for the jet of $3.6\times 10^{25}$~W for a distance of 10~kpc. Using the standard equations for particle and magnetic field energy \citep{1970ranp.book.....P} and assuming unity for the filling factor and ratio of proton to electron energies we find that the minimum energy field is 0.3~Gauss. The electron lifetime against synchrotron losses at 15~GHz in this field is $\sim1.7$~years whereas that for 1~keV X-rays is~$\sim1.5$~hours. Therefore electron energy losses suggest the radio should not decay quicker than the X-ray, if both emission mechanisms are synchrotron, contrary to what is observed in Figure~\ref{fig:ratio}.

These arguments suggest that whilst the jet maybe our principle protagonist for both radio and X-ray emission, their radiative mechanisms are very different. The radio emission is still almost certainly due to synchrotron radiation and $L_{\rm{Radio}}\propto Q_{\rm{jet}}^{17/12}$ (Equation~\ref{eq:jet_power}). The hard X-ray tail could be due to inverse Compton (or synchrotron self-Comptonization) from the jet, hence L$_{\rm{X-ray}}$ would be simply proportional to the outflow rate $\dot{m}_{\rm{jet}}$ and thus $\propto Q_{\rm{jet}}$, the jet power. The steady flux of X-rays would therefore originate from the base of the jet while the radio is emitted further down the jet. 
\rm
\section{Conclusions}

This work has shown, for the first time, a direct relationship between the X-rays and radio in the steady jet state of GRS\,1915+105. Previous attempts have failed to show this relationship~\citep[e.g.][]{2001ApJ...556..515M}, as they have included X-ray/radio comparisons that include extended knots and or X-ray accretion in other states.

\cite{2003MNRAS.344...60G} observed a universal radio-X-ray correlation in low/hard state black holes, but only for radiatively inefficient accretion. Figure~\ref{fig:universal_plane} shows the difference between the two models apply to other stellar-mass black holes and GRS\,1915+105. The difference between the two models is likely to be due to the rate of accretion or spin of the black holes. GRS\,1915+105 is known to be in a constant `soft'-like state as a large accretion rate is constantly present; however, it remains unclear if the X-ray emission, in the steady jet state, is produced from either the thin disk, an advection dominated flow or the compact jet. For other XRBs, the bright sources are likely to form only a transient soft-accretion disk and GRS\,1915+105-type accretion may only occur in systems with higher accretion rates, like AGN.

It is interesting that within each steady jet state, both the X-ray and radio luminosities fell with time. This suggests a cooling of the mechanisms or conditions that initially created the steady jet. Furthermore, whilst a coupling of the accretion process to the outflowing jet via advection is possible, the simplest explanation is that both the X-ray and radio emission originate directly from the jet as suggested by~\cite{2005ApJ...635.1203M}.

\begin{acknowledgements}
AR acknowledges support from an STFC studentship during this research and part of this work was also supported by the EXPReS project. Thanks are also given to Tom Maccarone and Elena Gallo for useful discussion throughout this work. EXPReS is an Integrated Infrastructure Initiative (I3), funded under the European Commission's Sixth Framework Programme (FP6), contract number 026642. The Ryle Telescope is operated by the University of Cambridge and supported by STFC. The X-ray data was provided by the ASM/\textit{RXTE} teams at MIT and at the \textit{RXTE} SOF and GOF at NASA's GSFC.
\end{acknowledgements}

\bibliography{references}
\bibliographystyle{aa}

\end{document}